\documentclass[a4paper,11pt]{article}
\pdfoutput=1 

\usepackage{jcappub} 
\newcommand{\apjl}{Astrophys. J. Lett.} 
\newcommand{\apjs}{Astrophys. J. Suppl. Ser.} 
\newcommand{\ao}{Appl. Opt.} 
\newcommand{\aap}{Astron. Astrophys.} 

\usepackage[T1]{fontenc} 
\newcommand{\mm}[1]{\mbox{$#1$}} 

\newcommand{\h}{\mm{h^{-1}}}

\newcommand{\bm}[1]{\boldsymbol{#1}}
\newcommand{\sbr}[1]{_{\textrm{#1}}} 
\newcommand{\inv}{\mm{^{-1}}}
\newcommand{\kmsmpc}{km s\inv Mpc\inv}
\newcommand{\ax}{\mm{a_x}}
\newcommand{\zcmb}{\mm{z_\text{CMB}}}

\newcommand{\zrec}{\mm{z_\text{rec}}}
\newcommand{\zcosmo}{\mm{z_\text{cosmo}}}
\newcommand{\Ho}{\mm{H_0}}

\newcommand{\LCDM}{\mm{\Lambda\text{CDM}}}
\newcommand{\mycomment}[1]{\ignorespaces}
\usepackage[sort&compress]{natbib}
\title{Uncertainties in the Hubble Constant from Peculiar Velocities}

\author[a,b]{Amber M. Hollinger}
\author[a,b,c]{and Michael J. Hudson}
\affiliation[a]{Department of Physics and Astronomy, University of Waterloo, 200 University Ave W, Waterloo, ON N2L 3G1, Canada}
\affiliation[b]{Waterloo Centre for Astrophysics, University of Waterloo, 200 University Ave W, Waterloo, ON N2L 3G1, Canada}
\affiliation[c]{Perimeter Institute for Theoretical Physics, 31 Caroline St. North, Waterloo, ON N2L 2Y5, Canada}

\emailAdd{amhollin@uwaterloo.ca}
\emailAdd{mike.hudson@uwaterloo.ca}

\abstract{
Recent measurements of the Hubble constant using type Ia supernovae explicitly correct for their estimated peculiar velocities using the 2M++ reconstruction of the local density field. The amount of uncertainty from this reconstruction procedure has thus far been unquantified. To rectify this, we use mock universe realisations of the 2M++ catalogue and generate predicted peculiar velocities using the same method as the predictions that are used to correct for the Pantheon+ catalogue. We find that the method yields uncertainties of  $\sim$ 0.3 \kmsmpc\ and hence subdominant to the total uncertainty in $H_0$.}

\arxivnumber{1234.5678}

\begin{document}
\maketitle
\flushbottom

\section{Introduction}\label{sec:intro}

The determination of the Hubble Constant (\Ho), which describes the current expansion rate of the Universe, provides a fundamental test of the standard cosmological model, $\Lambda$ Cold Dark Matter (\LCDM). Although highly successful in reproducing an abundance of observed cosmological phenomena, there is a significant disagreement in this model between independent early (`global') and late (`local') time measurements of \Ho, raising concerns about the reliability of the model. 
Comparing measurements at opposite ends of the observable expansion history of the Universe provides a rigorous test for evaluating the standard cosmological model.  Local measurements involve techniques such as the distance ladder, which extends to Type Ia Supernovae (SNe Ia) calibrated through geometric distances and Cepheid variables \citep[e.g.][]{riess_24_2016,reid_improved_2019}. Other local measurement methods include using the tip of the red giant branch (TRGB) \citep[e.g.,][]{freedman_measurements_2021,scolnic_cats_2023} or using gravitational waves as  standard siren measurements \citep[e.g.,][]{abbott_gravitational-wave_2021}. Assuming a standard cosmological model, global measurements of \Ho\ include observations of large-scale structure, using standard ruler methods (i.e. baryon acoustic oscillations or cosmic microwave background, hereafter CMB)\citep[e.g.,][]{pogosian_recombination-independent_2020, alam_completed_2021,planck_collaboration_planck_2020-1}.
Since the groundbreaking study by ref.\ \cite{riess_24_2016} (hereafter R16), which achieved a local determination of the Hubble parameter with an uncertainty of 2.4\%, there has been a growing tension between this value and the Planck value based on $\Lambda$CDM. This tension, known as the ``$H_0$ tension'', was initially significant at the $3.4 \sigma$ level (R16) but has steadily increased to now reach a statistical significance of $5.0 \sigma$ (\citep{riess_comprehensive_2022}, hereafter R22). 

Peculiar velocities are one of the factors that impacts distance-ladder measurements of $H_0$, because they perturb the cosmological redshifts, introducing uncertainty in determining the Hubble constant from low-redshift distance indicators.  In the peculiar velocity literature, it has long been known that one can make reasonably accurate predictions of these peculiar velocities using a galaxy density field and and linear perturbation theory \citep[see, for example, ref.][for a summary of work in this field by the mid-1990s]{strauss_density_1995}. In early studies of \Ho\ and of other cosmological parameters from SNe, however, these peculiar velocity corrections were ignored, although a number of studies quantified the uncertainty in \Ho\ due to the uncorrected peculiar velocities \citep{TurnerCenOstriker1992, MarraAmendolaSawicki2013, WojtakKnebeWatson2014, wu_sample_2018}. Ref.\ \cite{neill_peculiar_2007} explicitly showed their impact on the cosmological parameters $\Omega\sbr{m}$ and the dark-energy equation of state parameter, $w$. It is now standard practice to remove peculiar velocities, using a model of the peculiar velocity field \citep{willick_determination_2001, carrick_cosmological_2015, boruah_peculiar_2021, peterson_pantheon_2022}.
Hence accurate reconstructions of this peculiar velocity field from observed large-scale structures are essential for reliable inference of $H_0$. 

However, while the uncertainties in these reconstructions and consequently their impact on \Ho\ has been studied in an approximate way in previous works  \cite[such as][]{peterson_pantheon_2022}, the full impact of peculiar velocity reconstruction on \Ho\ has not been definitively quantified using a fully robust method. In this work we will provide a full assessment of the uncertainty introduced into \Ho\ due to the 2M++ velocity reconstructions. To achieve this, we use our mock 2M++ redshift compilation catalogues \cite[see ref.][hereafter HH24 for details]{Hollinger2024}  which  match the geometry and magnitude limits of the density field that was used to correct for the SNe peculiar velocities in R16 and R22. We do not use the actual 2M++ catalogue in this analysis as we are trying to determine the extent to which our predicted peculiar velocity methodology is able to reconstruct the true redshift. This is not possible without knowing the true peculiar velocity of the SNe. By using simulation data which matches surveys selection, we are able to determine the extent to which the reconstruction impacts measurements of \Ho\ for the first time. Hence we are able to more  fully understand the impacts these corrections take on the measurement of \Ho, focusing on the effect of peculiar velocities on the `distance ladder' measurement of R22.

This paper is organised as follows: in section \ref{sec:PVs+Ho} we describe the framework needed to generate peculiar velocities from a density field and the impact that both the density field and peculiar velocities have on measurements of \Ho. We introduce the supernovae catalogue used for this paper in section 3 and the methodology we adopt to calculate changes in \Ho. In section 4 we present our results and methodology in the reconstruction. Section \ref{sec:discussion} discusses potential limitations and caveats to this analysis and finally we present our conclusions in section \ref{sec:conclusions}.

Unless otherwise stated we assume the \LCDM\ parameters of the MDPL2 simulation \citep{klypin_multidark_2016} used later in this work, and described in more detail in section \ref{sec:sim_data}. These are based on \citep{planck_collaboration_planck_2014}: $\Ho = 67.77$ \kmsmpc, $\Omega\sbr{b}h^2 = 0.022139, \Omega\sbr{c}{h}^2 = 0.1189, n_s = 0.96, \sigma\sbr{8} = 0.8228$. 

\section{Peculiar Velocities and $H_0$}\label{sec:PVs+Ho}

The most recent calculations of the Hubble constant performed by R22 correct their final cosmological redshifts using the predicted peculiar velocities of ref.\ \cite{carrick_cosmological_2015} (hereafter, C15) and assign a generous 250 km/s uncertainty to each of these corrections. 

\subsection{Peculiar Velocities from the Density Field}

In the linear regime of perturbation theory,  it is possible to relate the peculiar velocity field to the underlying density field via: 
\begin{equation}\label{eqn:v_field}
      \boldsymbol{v}(\boldsymbol{r}) = \frac{H a f(\Omega_m)}{4 \pi}\int \delta(\boldsymbol{r}') \frac{ (\boldsymbol{r}'-\boldsymbol{r})}{|\boldsymbol{r}'-\boldsymbol{r}|^3} d^3\boldsymbol{r}' \, ,
\end{equation}
where $\delta=(\rho-\bar{\rho})/\bar{\rho}$ is the density contrast, and  $f \approx \Omega_m^{0.55}$ is the logarithmic growth rate of structure in \LCDM. Note that if $r$ is written in units of $h^{-1}$ Mpc, the dependence on $H$ drops out of this equation: no knowledge of $H$ is needed to predict peculiar velocities from a density field.

There are three caveats to this equation: 
\begin{enumerate}
    \item  Galaxy surveys are unable to directly measure the total matter density.  As a result it is typically assumed that there is a linear relation between the observed galaxy density field and the total matter density field, such that $\delta_g = b \delta$. This linear biasing term ($b$) is dependent on a number of factors, such as galaxy type, and is usually unknown. Replacing $\delta$ in eq.\ (\ref{eqn:v_field}) with $\delta\sbr{g}$ yields a prefactor $\beta \equiv f/b$.  However, $\beta$  can be measured directly by comparing a predicted peculiar velocity field to observed peculiar velocities (see refs. C15, \citep{boruah_cosmic_2020, lilow_constrained_2021} for recent examples), and this measurement is independent of $H$.
    
    \item The integral presented in eq.\ (\ref{eqn:v_field})\ is over all space, however galaxy surveys are finite due to instrument limitations. As such, our reconstructed density fields are insufficient to fully encapsulate the motions of galaxies, this is typically addressed by modelling the contribution from sources beyond the survey limits as a dipole, i.e.\ an ``external'' bulk flow, $\bold{V}_\text{ext}$. This term accounts for structure beyond survey limits and theoretically, with a large enough survey, decays to zero.
    
    \item Most importantly, the density fluctuations, $\delta\sbr{loc}=(\rho-\bar{\rho}\sbr{loc})/\bar{\rho}\sbr{loc}$, are normalised using the mean density  ($\bar{\rho}\sbr{loc}$) measured within a finite, local volume. This has important consequences if the local volume has a different mean density than the unknown global  $\bar{\rho}$, as is generally true due to cosmic variance. If we write  $ \frac{\bar{\rho}\sbr{loc}}{\bar{\rho}} \equiv 1 + \Delta$, then 
    $\delta = \delta\sbr{loc} (1+\Delta) + \Delta$\,. 
    Similarly for the local fluctuation of galaxy (as opposed to matter) density we have $\delta\sbr{g} = \delta\sbr{g,loc} (1+\Delta\sbr{g}) + \Delta\sbr{g}$, where, assuming linear biasing, $\Delta\sbr{g} = b \Delta$.
    
\end{enumerate}

Putting all of these together, the model for the predicted peculiar velocity becomes 
\begin{equation}\label{eqn:v_pred}
      \boldsymbol{v}\sbr{pred,c}(\boldsymbol{r}) = \beta(1+\Delta\sbr{g}) \left[\frac{H a}{4 \pi}\int_{V\sbr{surv}} \delta\sbr{g,loc}(\boldsymbol{r}') \frac{ (\boldsymbol{r}'-\boldsymbol{r})}{|\boldsymbol{r}'-\boldsymbol{r}|^3} d^3\boldsymbol{r}' \right] + \bold{V}_\text{ext} - \frac{\beta\, H a}{3} \Delta\sbr{g} \boldsymbol{r} \,,
\end{equation}
where the integral is now over the finite survey volume $V\sbr{surv}$ and the last term comes from applying Gauss' law. 
In practice, density field reconstruction combined with observed peculiar velocity data, allows for the determination of the velocity scale parameter $\beta$ and residual bulk velocity $\bold{V}_\text{ext}$ as follows. The reconstruction process generates an \emph{unnormalised} peculiar velocity field, $\boldsymbol{v}_\text{pred}(\bold{r})$ (i.e. the term in square brackets) and hence predicts the directions and magnitudes of peculiar velocities at different positions. These predictions are compared to galaxies with observed peculiar velocities, based on distance indicators such as Type Ia SNe, the Tully-Fisher or Fundamental Plane relations, to fit for $\beta$ and residual bulk velocity $\bold{V}_\text{ext}$.

The term $\Delta\sbr{g}$ is difficult to measure and is typically either marginalised over and then ignored, or assumed to be zero. Assuming the Copernican principle, that we don't live in a special place in the Universe, the expectation for the mean $\langle \Delta\sbr{g} \rangle = 0$, consistent with the above assumption. But  the standard deviation of this quantity, i.e.\ its cosmic variance, is estimated to be approximately 3\% for the 2M++ volume used here. The effect of cosmic variance on determinations of $\beta$ (and hence $f\sigma_8$) was discussed in Refs.\ \citep{hollinger_assessing_2021} and HH24. The effect of this cosmic variance on determinations of \Ho\ is the topic of this paper. 

\subsection{Reconstructing Cosmological Redshifts}\label{sec:recon}

At low redshifts  ($z<0.1$) the uncertainties in an object's cosmological redshift (\zcosmo) are primarily attributed to two sources: uncertainties in measurements and peculiar velocities.  In practice, however, redshift measurement errors are much smaller than the uncertainties in the peculiar velocity corrections. 
This is because  \zcosmo cannot be measured directly, rather it must be inferred from the observed redshift after it has been corrected to be in the CMB rest-frame (\zcmb), using the peculiar redshift approximated by $z_\text{pec} = v_\text{pec}/c$, where $v\sbr{pec}$ is the peculiar velocity, also in the CMB frame. The non-relativistic Doppler shift approximation is valid here since peculiar velocities are typically a few hundred km/s and always $v_\text{pec} \ll c$. The two quantities can then be related using:
\begin{equation}\label{eqn:zcosmo_from_zcmb_zpec}
    z_\text{cosmo} = \frac{1+z_\text{CMB}}{1+z_\text{pec}}-1 \, .
\end{equation}
We define our reconstructed cosmological redshifts (\zrec) as 
\begin{equation}\label{eqn:zcosmo_from_zcmb_zpec}
    z_\text{rec} = \frac{(1+z_\text{cosmo})(1+z_\text{pec})}{(1+z_\text{pred,c})}-1 \,,
\end{equation}
where
\begin{equation}\label{eqn:zcosmo_from_zcmb_zpec}
    z_\text{pred,c} \approx \frac{v_\text{pred,c}}{c} = \frac{\beta v_\text{pred}+ \bold{V}_\text{ext}\cdot\bold{\hat{r}}}{c}\, .
\end{equation}

Various efforts have gone into trying to understand the influence of systematic uncertainties in peculiar velocity measurements and are considered in analyses measuring and $w$ \citep{scolnic_complete_2018,brout_first_2019} and \Ho\ with SNe. 
R16 examined the impact of peculiar velocity corrections on $H_0$ measurements by comparing a minimal redshift cut of $z = 0.01$ against that of $z = 0.023$ for their sample. They found that using the latter cut reduced sensitivity to peculiar velocities, finding a difference in the recovered value of $H_0$ of $0.3$ \kmsmpc. Ref.   
\cite{peterson_pantheon_2022} investigated the impact of large-scale coherent motions and small-scale virial motions of galaxies within groups, correcting for the peculiar motion of both, impacted their measure of $H_0$ by 0.06 - 0.11 \kmsmpc. However to date, the impact of peculiar velocity reconstruction and its application to the SNe in the Riess et al.\ papers has not been accurately quantified.

\subsection{Effects of an Under-Dense Local Universe}

There have been several observations, performed at multiple wavelengths, which find supporting evidence that there is a large local under-density around the Local Group. Initially it was observed in optical samples as a deficiency in the galaxy luminosity density \citep{maddox_galaxy_1990}, which  estimated that the local under-density spanned a region out to a distance of $\sim 140$ \h Mpc 
\citep{zucca_eso_1997}.  X-ray galaxy cluster surveys such as REFLEX II \citep{bohringer_extended_2015} and CLASSIX \citep{bohringer_observational_2020} also find an apparent local Universe  under-density of $\sim 30\%$ on similar scales. 
Additionally multiple studies focusing on data from near-infrared surveys have found that the under-density 
extends to even larger radii measuring it to scales of 200-300 \h Mpc \citep{frith_local_2003,busswell_local_2004,frith_new_2006, keenan_evidence_2013, whitbourn_local_2014,hoscheit_kbc_2018,haslbauer_kbc_2020,wong_local_2022}. 
However, whether or not we exist within a local under-density is still controversial. Ref. \cite{jasche_physical_2019}, using the 2M++ Catalogue, studied the distribution of galaxies and  found that local structures could be explained within the $\Lambda$CDM framework and found no evidence for an under-density on the proposed scales. 

Regardless of its existence, a spherical top-hat perturbation in our local matter density field, to first order, at low redshift ($a=1$) would generate a change in the local expansion rate as follows: 
\begin{equation}\label{eqn:deltaH0}
    \frac{ H\sbr{loc} - H_0}{H_0} = -\frac{1}{3} f(\Omega_m) \Delta
\end{equation}
\citep{turner_relation_1992},
where here 
$\Delta$ is the \emph{local} density contrast. From this it is clear that a local under-density, $\overline{\delta} < 0$ would result in local expansion rate which is higher $\Delta H_0/ H_0>0$. However, as we consider increasingly larger scales, these perturbations become smaller and their amplitude, which is constrained by observations of measurements of the power spectrum of matter density fluctuations, decreases. This results in a decrease in the density contrast which then tends to the homogeneous FLRW limit. It has been suggested that this cosmic variance on $H_0$ results in a systematic uncertainty of about $\sim$ 1\% when analysing observations within the redshift range of $0.023 < z < 0.15$ \citep[for more details see][and references therein]{camarena_impact_2018,odderskov_variance_2017}. 
Despite this uncertainty, it falls short of explaining why there is a significant difference of around 8\% between early-time and late-time $H_0$ constraints, which is impossible in $\Lambda$CDM.

Additionally ref. \cite{wu_sample_2018}, hereafter WH18, investigated if sample variance in the local Universe could alleviate the Hubble tension. This was done using N-body simulations to model local measurements and quantified the variance as a result of these local Universe density fluctuations and sample selection. While they directly matched the cosmological redshifts of the Pan-STARRS SNe in their model, they did not account for host galaxies peculiar velocities. They estimated the local uncertainty of $H_0$ due to sample variance to be 0.31 \kmsmpc.
Ref. \cite{kenworthy_local_2019} find a consistent result with predicted cosmic variance of WH18. They used a SNe sample consisting of data from  Pantheon, Foundation, and the Carnegie Supernova Project, and searched for evidence of a local void via the presence of large-scale outflows. They found no evidence to suggest the existence of a local large-scale under-density within $0.023<z<0.15$ with a high confidence of $\sim 4-5 \sigma$. 

The R16 and R22 papers explicitly correct for each SN's peculiar velocity using the reconstruction of the local 2M++ density field from C15. This correction is normalised within the survey volume of 2M++. As a result if the entire region 2M++ region is under-dense, there would be a net change in the peculiar velocities, that is indistinguishable from a change in the Hubble constant, but this would not included in the 2M++ predictions. Additionally, the reconstruction of the density field is subject to wide range of systematic errors (for details refer to HH24), hence still might require a partial correction for the sample variance.

\subsection{Constructing a Hubble Diagram}\label{sec:constructing}

The R22 measurement of the Hubble parameter is based on SNe Ia calibrated using a 3-rung distance ladder, see section \ref{sec:Pantheon} for more details. 
However, to estimate distances to SNe Ia one must assume that an object's absolute luminosity is the same as its local counterpart measurements, after applying corrections, the comparison of these values provides a measure of the distance modulus:
\begin{equation}\label{eqn:mu1}
   \mu_{x}^0 = m_x^0 - M^0_x = 5 \log_{10} \left(\frac{D_L(z)}{1 \text{Mpc}} \right)+ 25 \, ,
\end{equation}
where $x$ denotes the wavelength band and the superscript $0$ denotes a magnitude corrected for (or free of) any intervening absorption due to interstellar dust. One can also compute the luminosity distance $D_L$ of a light source with redshift $z$ in the context of General Relativity. Assuming a flat FLRW metric, one finds
\begin{equation} \label{dl}
    D_L(z) = (1+z) c \int^z_0 \frac{dz}{H(z)} \, ,
\end{equation}
where $c$ is the speed of light and $H(z)$ is the Hubble function. 

The luminosity distance at relatively low redshifts ($z \ll 1$), can be usefully expressed as a series expansion approximation in redshift \citep{riess_redetermination_2009}, resulting in
\begin{equation}\label{eqn:dl2}
\begin{aligned}
    D_L(z) &\approx \frac{cz}{H_0}\left[ 1 + \frac{z}{2}(1-q_0) - \frac{z^2}{6}(1-q_0-3q_0^2+j_0) + \mathcal{O}(z^3) \right] \, ,
\end{aligned}
\end{equation}
where  $q_0=-0.55$ and $j_0=1$ are the present deceleration and jerk parameters.

Equating eqs. (\ref{eqn:mu1}) and (\ref{eqn:dl2})  and introducing a constant $a_x$ defined to be the intercept of the magnitude-redshift relation:
\begin{equation}\label{eqn:ax}
    a_x = 
    \log D_L(z)
    - \frac{m_x^0}{5} \, .
\end{equation}
We emphasise that this equation depends on the redshift used: ideally one would use the true \zcosmo\ of the SNe, but in practice the redshifts used are either \zcmb \  or \zrec.
We can obtain an estimate of the Hubble constant that is solely reliant on two fitted parameters:
\begin{equation}\label{eqn:logH0}
    \log H_0 = 0.2\,M_x^0 + a_x + 5 \, .
\end{equation}

It is obvious from eqs. (\ref{eqn:dl2})-(\ref{eqn:logH0}) that the overall uncertainty in $H_0$ depends on (1) the absolute calibration $M_x^0$, (2) uncertainties in the apparent magnitudes of SNe and (3) the SNe's cosmological redshifts. The intercept $a_x$ is determined from a Hubble diagram of SNe Ia where the individual measurements of $m_x^0$ are determined with a light curve fitter \citep{scolnic_pantheon_2022}. 
While contributions in the overall uncertainty due to corrections performed (1) and (2) on the SNe's magnitudes have been explored in works \citep[e.g.,][]{carr_pantheon_2022,peterson_pantheon_2022}, we limit our focus in this work to the effects of (3).

For example, suppose our local universe was uniformly under-dense by 30\% out to a distance of 200 $h^{-1}$ Mpc from the Local Group. This would lead to a $\sim 5$\% fluctuation in the Hubble constant, if all SN used in the determination of \Ho\ were within that radius. Moreover, for this case, 2M++ would not correct for the under-density because 2M++ is wholly contained within the under-density. Such an under-density would reduce the tension between the local and CMB measurement of $H_0$ to $1.6\sigma$. However, such a fluctuation is very unlikely in \LCDM, as we will now show.

We can estimate the root mean square (r.m.s.) for such density fluctuations in \LCDM\ by noting that the r.m.s.\ mean density for a sphere of radius R is given by 
\begin{equation}\label{eqn:sigmaR}
   \langle \delta^2_R\rangle \equiv \sigma^2(R) = \frac{1}{2\pi^2}\int P(k) \, \tilde{W}(k,R)^2 \, k^2 \, dk \ ,
\end{equation}
where $\Tilde{W}(k,R)$ is the Fourier transform of the spherical top-hat window function,
\begin{equation}
    \tilde{W}(k,R) =  \frac{3}{(kR)^2} \left[\sin(k R) -kR \cos(k R) \right] .
\end{equation}
The standard deviation in $\overline{\delta}$ can then be directly converted to a fluctuation in \Ho\ on the same scale using eq.\ (\ref{eqn:deltaH0}). Applying this to the scenario sketched above, we find that the r.m.s.\ halo mass density fluctuation in a 200 $h^{-1}$ Mpc sphere is 0.025. A 30\% under-density would therefore represent a 12$\sigma$ fluctuation, which is exceedingly unlikely in \LCDM.

\section{Cosmic Variance in $H_0$ with No Peculiar Velocity Reconstruction}

Before discussing the case where we use mock 2M++ reconstructions to predict the peculiar velocities, we first consider the cosmic variance for cases without any reconstruction. Understanding this will be necessary even for the 2M++ peculiar velocity reconstructions because these are only valid to $z \sim 0.07$, and so we will need to account for cosmic variance on scales $z \gtrsim 0.07$. In this section, we use a semi-analytic prescription to estimate the cosmic variance and compare it with previous results. In section \ref{sec:fullanalysis} we will also compare this with numerical simulations.


\subsection{Pantheon+ Catalogue}\label{sec:Pantheon}
In this paper, we want to emulate the analysis of the Pantheon+ catalogue, including the weighting and fitting method used in the determination of \Ho\ in R22. 
The Pantheon+ catalogue is the successor to the original Pantheon SN sample, and its analysis was a joint effort with the SH0ES team \citep{carr_pantheon_2022}. Pantheon+ consists of 1701 measurements of 1550 SNe between $0.001 < z < 2.26$.
Here we briefly review how the Hubble constant was measured in R22. The R22 measurement of the local $H_0$ uses distance anchors which have geometric measurements and hence precise distances. These anchors include independent calibration of Cepheids in NGC 4258 
, the Milky Way 
and the Large Magellanic Cloud. 
By using these anchors to calibrate $\mu^0$ of host SNe galaxies.  The distance ladder's second rung is composed of a set of 42 calibrators containing Cepheids and SNe; with the anchors setting the Cepheids' absolute distance. Within the distance ladder's third rung of Pantheon+, there is a subset of 277 SNe (located within the Hubble flow (HF) at $0.023<z<0.15$ limited to but not inclusive of all SNe within this redshift range) which we will refer to in this work as the SH0ES HF sample. The global fit of all these rungs, is the set of standard candles that are used to determine a measurement of \Ho.

\subsection{Perturbing Cosmological Redshifts with Density Fluctuations}\label{sec:perturbing}

In this section we aim to simulate the impact that cosmic variance introduces into the derived Hubble constant due to uncorrected peculiar velocities. In HH24, we were able to generate 15 independent local universes to match the boundary conditions of the 2M++ survey, of diameter 400\h Mpc from the 1 \h Gpc MDPL2 box, but we are unable to extend these boundaries and maintain independence. In order to maintain their independence we instead draw density fluctuations for nested spheres from a Gaussian distribution to simulate velocity perturbations.   
We do this using the assumption that monopole fluctuations in the density field are the dominant source of cosmic variance in \Ho. We will justify this with full N-body simulations later in section \ref{sec:monopole}.

In section \ref{sec:constructing}, we discussed the case of a single under-dense region.
In a more realistic case, we have nested spheres of radii $R_1, R_2$ with their covariance being given by a variation of eq.\ (\ref{eqn:sigmaR}), namely:
\begin{equation}\label{eqn:covar}
\sigma_{R_1,R_2} = \frac{1}{2\pi^2}\int P(k)\tilde{W}(k,R_1)\tilde{W}(k,R_2) k^2 dk \ .
\end{equation}
We perform our power spectrum calculations using the publicly available python package ``Code for Anisotropies in the Microwave Background'' \citep[CAMB;][]{Lewis_camb_2000}. In doing so we assume the \LCDM \ parameters given in section \ref{sec:intro}.

We simulate peculiar velocities from these nested spheres as follows. First, we draw a sample of 250 spherical shell radii in redshifts that are evenly spaced on a log scale for the full range of Pantheon+ redshifts.
We then calculate the change in velocity ($v_\text{var}$) that would be associated for a SN located in that sphere, at its given comoving radius: 
\begin{equation}\label{eqn:vvar}
    v_\text{var} = -\frac{100}{3} 
    f(\Omega\sbr{m,0})
    \delta\sbr{n}(R) R \,,
\end{equation}
where the nested density fluctuations $\delta\sbr{n}(R)$ are drawn from a multivariate Gaussian with a mean of zero but whose covariance is given by eq. (\ref{eqn:covar}). 
Thus for each mock SN catalogue, we have generated peculiar velocities ($v\sbr{var}$) which are then used to generate mock ``CMB frame'' peculiar velocities as follows: $z_\text{fit} = (1+\zcosmo)(1+ v_\text{var}/c)$. 

\begin{figure*}
    \centering
    \includegraphics[width=\textwidth]{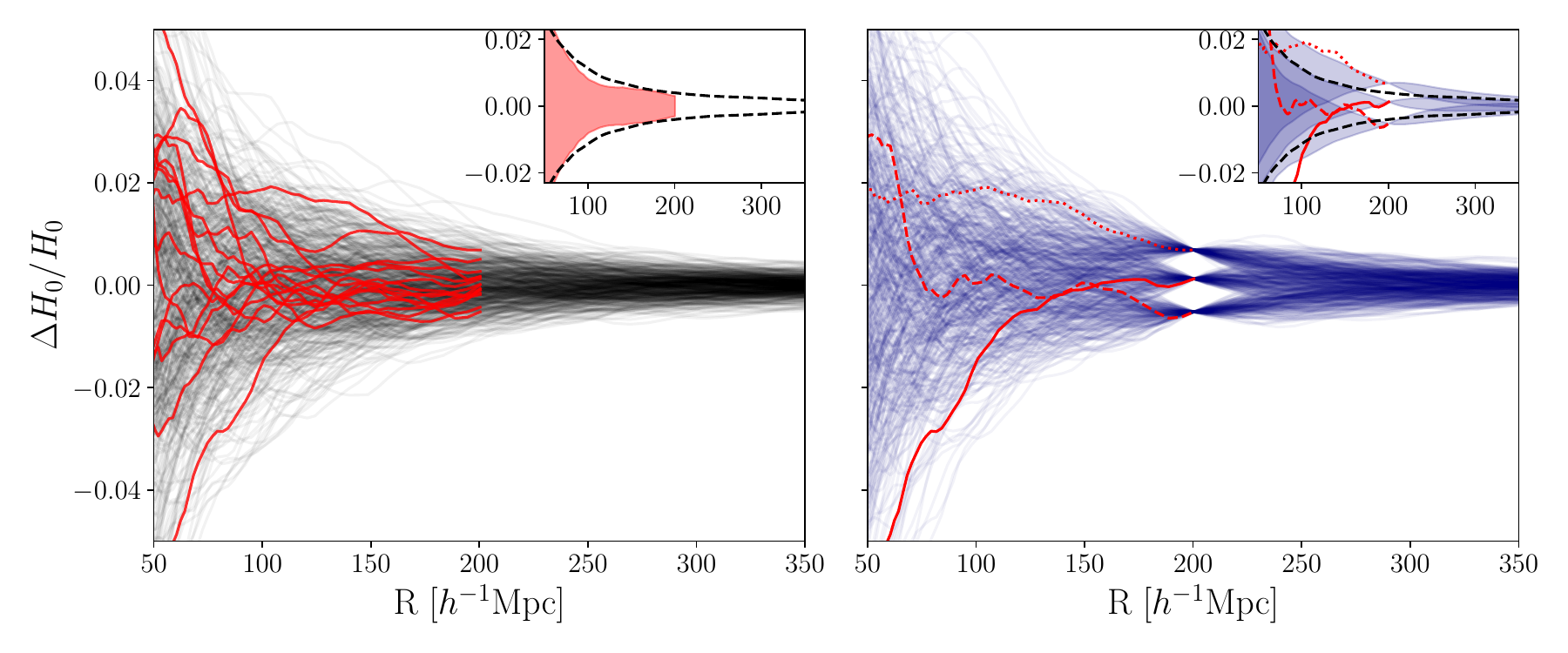}
    \caption[test]{The change in \Ho\ as a function of distance due to perturbations in the local matter density field. The left (right) panel shows 500 (167) density perturbation realizations drawn from a multivariate Gaussian matching for one (three) of our mocks that are plausible within the constraints of \LCDM. The left panel has no additional constraints, overlaid in red are the measured values of $\Delta_\textrm{particles}$ from each of the 15 mocks. The realizations shown in the right panel have the additional constraint that the $\Delta_\textrm{particles}$ at a radius of 200 \h Mpc matches that the measured from the mocks, which are shown in red. 
    The black dashed lines on the inlaid panels shows the standard deviation (std.) of the unconstrained realizations. The std.\ from the 15 particle densities are shown on the left in red, while the constrained std.\ of the lowest, mean, and highest perturbations at 200 \h Mpc are shown in blue on the right. 
    }
    \label{fig:deltaHo}
\end{figure*}

Shown in figure \ref{fig:deltaHo} is the expected change in the measurements of \Ho\ as given by eq. (\ref{eqn:deltaH0}), due to local density perturbations. The right panel show selected realizations that are constrained at 200 \h Mpc by the density contrast as measured by the MDPL2 particle catalogues. 
As demonstrated in the inlaid panel, the standard deviation of density fluctuations for the particles closely matches the \LCDM\ expectations. 

\subsection{Calculating Changes in \Ho}\label{sec:ax}

In the previous section, we calculated mock ``CMB frame'' redshifts. We can now investigate the impact this has on the recovered value of $a_x$ and hence \Ho.  For each mock catalogue we perform a weighted fit for $a_x$ using the statistical and systematic covariance matrix provided by \citep{brout_pantheon_2022}. We begin by reducing the full covariance matrix into one that only includes the SNe in the redshift range that we are investigating. 
We then perform a simple $\chi^2$ minimisation procedure to determine the best value of $a_x$:
\begin{equation}
    \chi^2(a_x) = \bold{\Delta}^T C^{-1} \bold{\Delta} \, ,
\end{equation}
where $\Delta$ is a vector of residuals of eq.\ (\ref{eqn:ax}). %
By comparing how these measurements of \ax\ change with respect to the base measurement, we can then propagate the standard deviation in \ax\ ($\sigma_{\ax}$) into the standard deviation of \Ho\ via:
\begin{equation}
    \sigma_{\Ho} =  \ln (10)\, \Ho\, \sigma_{\ax} \,.
\end{equation}
We note that for the purposes of this calculation we assume the R22 value for \Ho\ of 73.04 \kmsmpc.

As in the case of the true 2M++ density field, the peculiar velocity corrections are not perfect. The goal of this work is to determine how much uncertainty gets propagated due errors in the reconstruction of the peculiar velocities and uncertainties in the bulk flow. In this work we assume that the B-band magnitudes and the `Hubble Diagram' (cosmological) redshifts of R22 to be entirely accurate, to be used as a consistent reference point. We keep consistent weighting throughout the entirety of this work using the covariance matrix presented in R22, as the choice of weights would also impact the final recovered value of the intercept \ax. We are interested in the difference in \ax\ measured from \zcosmo\ and that obtained using \zrec, and its scatter from mock-to-mock. Additionally throughout this work unless stated explicitly we ignore contributions to \ax\ from objects classified as calibrators within the Pantheon+ catalogue, their influence will be expanded upon in section \ref{sec:calibrators}.

\subsection{Density Fluctuations impact on \Ho}\label{sec:monopole}

As discussed in section \ref{sec:perturbing}, we assume the variations in the Hubble flow come predominantly from the monopole, the mean density fluctuation.  We use a randomly generated artificial density perturbation, which in turn generates local variations in the expansion rate of the universe. Thus for each SN in the Pantheon+ catalogue we vary the redshift via eq. (\ref{eqn:vvar}), and using the method described in section \ref{sec:ax} calculate the change in the global measurement of the Hubble constant.

We generate 500 mock SN samples replacing the true $z\sbr{cosmo}$ with $z\sbr{fit}$ as described above. In doing so, we find that the unconstrained density fluctuations generate a scatter of $0.63 \pm 0.02$ \kmsmpc\ in \Ho. Repeating the exercise for the 217 SN from the Supercal HF sample studied by WH18, we find $0.41 \pm 0.01$ \kmsmpc. This agrees marginally with with the WH18 results who estimated 0.31 \kmsmpc\ from N-body simulations. We note that in this section calculations to the change in \Ho\ are performed solely using the diagonal of the matrix to match the WH18 methodology, the analyses in the remainder of this paper will be performed using the full covariance matrix. 

\section{Cosmic Variance in \Ho\ with mock 2M++ predicted peculiar velocities}\label{sec:fullanalysis}

Now we turn to the main focus of this paper, which is estimating how the uncertainties in 2M++ peculiar velocities affect the uncertainties in \Ho. We will do this by using mock 2M++ catalogues drawn from N-body simulations, reconstructing the peculiar velocities from these simulations and comparing the fitted value of \Ho\ using these corrections with the true value in each mock.

\subsection{Simulation Data}\label{sec:sim_data}

In this analysis we use data from the publicly available MultiDark Planck 2 simulation (MDPL2), which is one of a suite of cosmological gravity-only $N$-body simulations that are a part of the MultiDark project \citep{klypin_multidark_2016}. This simulation consists of $3840^3$ particles evolving in a periodic box of cubic volume of length 1 \h Gpc starting from a redshift of $z=120$ to $z=0$ with a mass resolution of $m_p = 1.51 \times 10^9 \h M_\odot$.  The adopted cosmology follows that of a flat \LCDM\ model and assumes the Planck parameters: $\Omega_\Lambda=0.692885$, $\Omega_M=0.307115$, $h=0.6777$,  $\sigma\sbr{8,linear}=0.8228$, and $n_s=0.96$ as per ref. \cite{planck_collaboration_planck_2014}. Haloes and subhaloes are identified using \textsc{rockstar} \citep{behroozi_rockstar_2012}, while merger trees are constructed using \textsc{consistent trees} \citep{behroozi_gravitationally_2013}.

Our analysis uses the Semi-Analytic Galaxies (SAG) catalogue of galaxies within MDPL2, which is generated via a post-processing step which places galaxies onto $N$-body simulations. This semi-analytic model (SAM) approach is significantly more computationally efficient when compared to full hydrodynamical simulations with self-consistent baryonic physics. The baryonic prescription of SAG is based on the work of ref. \cite{cora_semi-analytic_2018}, this incorporates crucial physical processes in galaxy evolution and formation, such as galaxy mergers, radiative cooling, chemical enrichment, supernova feedback and winds, star formation including starbursts and disc instabilities. For a detailed overview of this model, or the other SAMs used in MDPL2, refer to ref. \cite{knebe_multidark-galaxies_2018}. We use the snapshots taken at $z=0$ for both the halo and galaxy catalogues, are publicly available through the MultiDark Database \citep{riebe_multidark_2013} and can be obtained from the \texttt{CosmoSim} database \footnote{\url{www.cosmosim.org}}.

\subsection{Mock 2M++ Catalogues and Peculiar Velocities}

The peculiar velocities generated using 2M++ have been used by numerous other works. In HH24, we determined the extent to which previous measurements of $f\sigma_8$ were impacted by the selection biases inherent in 2M++. This was accomplished by generating mock catalogues of 2M++ using MDPL2 and its SAMs to recreate the conditions that would be introduced. For a detailed description of how this was done, refer to HH24.  In brief,  
\begin{enumerate}
    \item We cut the simulation down to 15 non-overlapping spheres to a  maximum survey depth of 200 \h Mpc ($K$<12.5), to be consistent with the hard radius cuts applied in the analysis of C15. 
    For each sphere, we take the Cartesian velocities and project them to a radial velocity. 
    \item We used abundance matching to generate the $K$-band luminosity of the galaxies from the provided SAM stellar masses in each one of these local universes, using the Schechter function parameters of C15.
    \item We then determine if each galaxy could be observed from the centre of a mock universe, by calculating the galaxies' effective apparent magnitudes. Galactic latitude and flux limit cuts are then applied to replicate the observational constraints of the 2M++ survey.
    \item We generate the density field by first placing the galaxies onto a padded three-dimensional grid. As there are several flux limits for 2M++, the galaxies are then weighted according to the luminosity weighting scheme described in HH24. As the density field encompasses all observed objects within the given flux limits, it is populated with both central and satellite galaxies. Subsequently, we derive the density perturbation field, $\delta\sbr{t}(\boldsymbol{r})$, and apply a Gaussian smoothing kernel of 4 \h Mpc in Fourier space, which was chosen to be consistent with C15, \citep{hollinger_assessing_2021}, and HH24. 
    \item From the smoothed density perturbation field, the velocity field can be calculated easily in Fourier space via the Fourier transform of eq. (\ref{eqn:v_field}): 
    \begin{equation}
    \bm{v_k}=i H a f \frac{\bm{k}}{|\bm{k}|^2}\delta_k \\.
    \end{equation}
    This provides the predicted peculiar velocity field in Cartesian coordinates, and subsequently galaxies are assigned a radial peculiar velocity. We note that for our purposes, $\bm{k}$ is expressed in units of $h$ Mpc$^{-1}$, and that we set $H_0 = 100 h$ km s$^{-1}$ Mpc$^{-1}$, as the factors of $h$ is cancel, which renders our predicted velocity to be independent of any choice of \Ho.
\end{enumerate}

\begin{figure*}
    \centering
    \includegraphics[width=\textwidth]{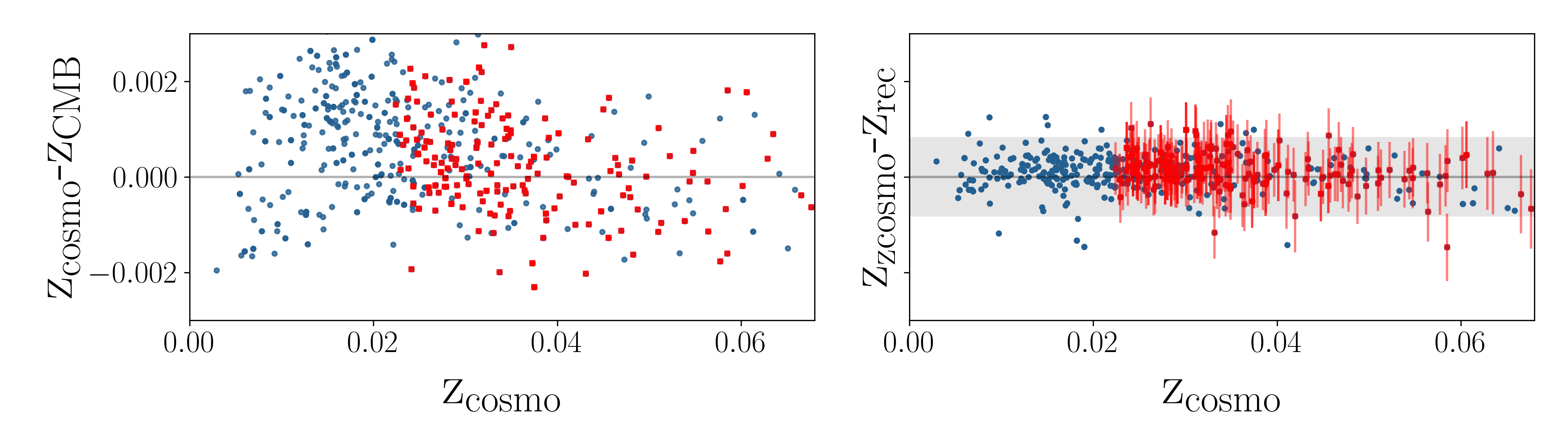}
    \caption[test]{Left panel: the difference between the `observed' CMB frame redshift \zcmb\ and the cosmological redshift (\zcosmo). These are the 601 SNe that lie within the survey limits of 2M++, the 207 SN in the HF sample are distinguished by red square markers. Right panel:  the difference between reconstructed redshifts after accounting for external flows, peculiar velocities and galaxy bias (\zrec) and \zcosmo, for one of our mocks. For comparison the 250 km s\inv\ assumed in R22 is show in the grey band while the 2M++ expected predicted peculiar velocity uncertainty in our SNe mocks (120-260 km/s) calculated in HH24 is shown by the error-bars.}
    \label{fig:zcosmo_zrec}
\end{figure*}

\begin{figure*}
    \centering
    \includegraphics[width=.9\textwidth]{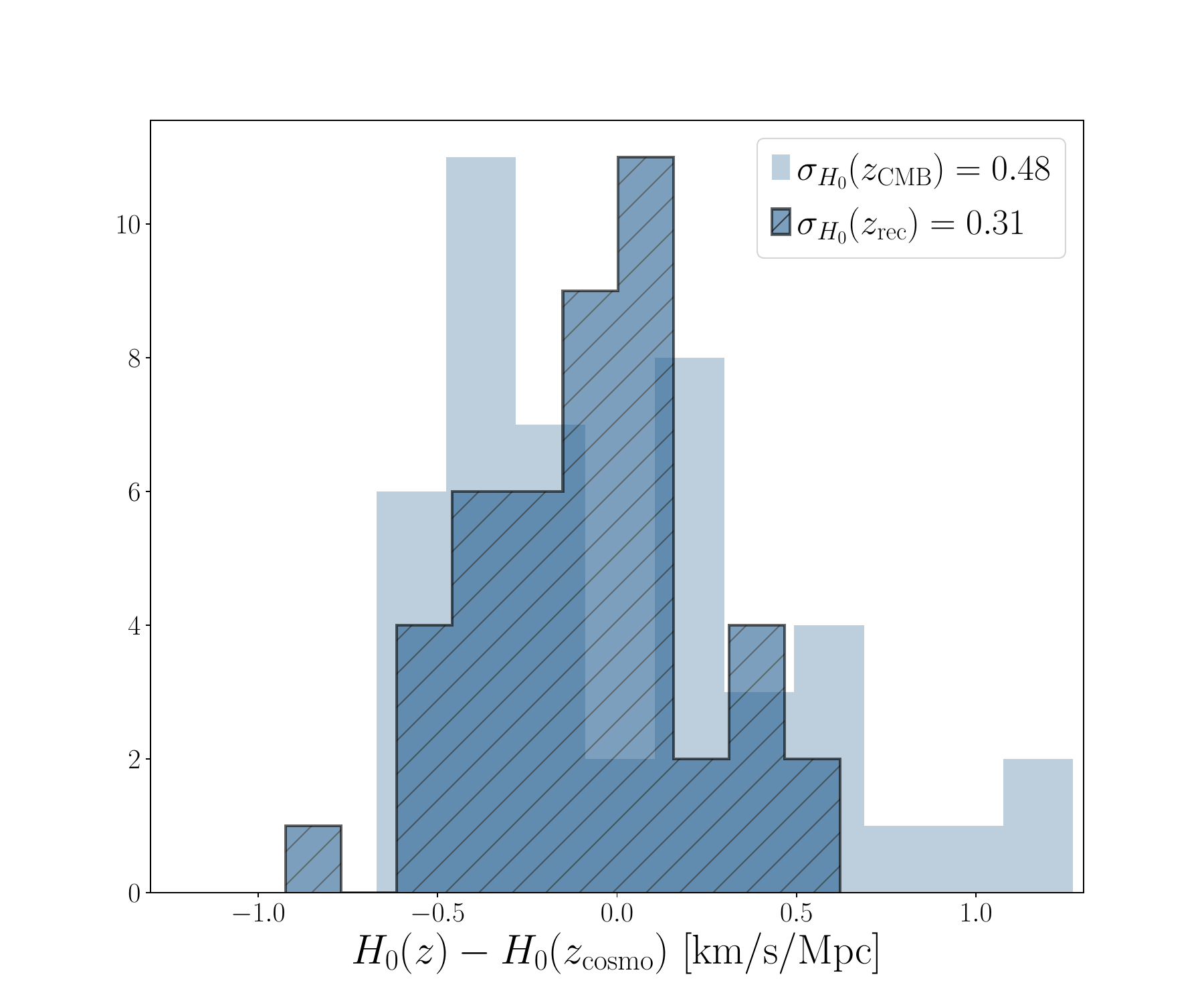}
    \caption[ ]{Histograms of the error in \Ho, i.e.\ the difference between \Ho\ calculated using the  `observed' CMB frame redshifts (unhatched) or reconstructed redshifts (hatched) against the true \Ho\ using the exact cosmological redshifts. In both cases, the errors are based on SH0ES HF SN sample for all 45 mock universes. The values in the legend are the standard deviation of each sample, in units of \kmsmpc.}
    \label{fig:axcomp}
\end{figure*}
\subsection{Redshift Reconstruction using Mock 2M++ Catalogues}

The Pantheon+ data used in  R22 provides a measure of objects' Hubble-diagram redshift, which is the peculiar velocity corrected CMB frame redshift.  To avoid confusion we will refer to our mock galaxy catalogue redshifts as follows: \zcosmo, the true cosmological redshift based on the comoving coordinates of our mock catalogues;  \zcmb, the ``observed'' CMB-frame redshifts after correcting \zcosmo\ by the true peculiar velocities; 
and \zrec\ the reconstruction of the cosmological redshifts from  \zcmb\ after correcting for the predicted peculiar velocities' measured bulk flow and $\beta$. 

As we are working with mock 2M++ catalogues, the physical locations of the galaxy hosts in the Pantheon+ catalogue do not have precise matches in the mocks. Our procedure for generating our mock Pantheon+ samples  can be summarised as follows: 
\begin{enumerate}
    \item Each Pantheon+ SN located within the volume of the 2M++ catalogue is matched to the nearest proxy central galaxy from our HH24 catalogues. Any galaxy located outside the confines remains at its given Hubble Diagram redshift, we refer to these updated redshifts as \zcosmo. 

   \item We alter \zcosmo\ by adjusting the final redshifts of our mock catalogues used in our calculations. The method used to alter the redshift depends on whether the SN lies inside or outside the 2M++ volume.
   
   \begin{enumerate}
   
        \item If the object is located \emph{within the volume of 2M++}: we alter \zcosmo\ using the description outlined in section \ref{sec:recon} using both the true N-Body and predicted radial peculiar velocities from our catalogues. Note: 
        While the R22 calibrator objects lie with the 2M++ volume we do not use them in the majority of our analysis, see section \ref{sec:calibrators}.
   
        \item If the object lays outside of the survey volume its redshift is instead corrected by an artificial density fluctuation to maintain the independence of the 15 mock universes, using the method described in section \ref{sec:perturbing} with an added constraint, the density fluctuations chosen for this sample are not arbitrary. Instead they are constrained by imposing the requirement that the density fluctuations at a radius of 200 \h Mpc must match that of that local universe's over or under density from MDPL2's true particle density, see figure \ref{fig:deltaHo}. 

    \end{enumerate}
    We refer to the set of these altered redshifts as \zrec.
    \end{enumerate}

Shown in figure \ref{fig:zcosmo_zrec} is the redshift difference between the CMB frame (left) and our reconstructed redshifts (right) against the true cosmological redshift of the SN within the survey limits of 2M++ . This demonstrates the impact of peculiar velocities on observations and the extent to which we are are able to correct for them using our reconstruction technique.  The scatter generated solely as a result of peculiar velocities for our mocks is on average 275 km s\inv, while, for the reconstruction, this is reduced to 132 km s\inv. The latter scatter is slightly less than the $\sigma_v \sim 170$ km/s found in HH24. The gray band shown on the bottom right panel of this figure is the 250 km s\inv\ uncertainty that is assumed for the R22 corrections. For visualisation purposes only, we show the predicted uncertainties from our results in HH24, which ranges between 120-260 km s\inv.

\subsection{Reconstruction Variations}

We can now perform the same calculation described in section \ref{sec:ax} through the comparison of measurements made between \zrec\ and \zcosmo. However, before doing so we need to acknowledge that our calculated values have some additional sources of uncertainty which may result in our estimates that peculiar velocities introduce into the \Ho\ are a little too conservative. To compensate for this we list here some of the potential sources and how these impact the final results listed in table \ref{table:HOcomp}. 
\begin{itemize}
    \item The measured fluctuations in mean density for our halo weighted samples are based on our mocks being perfectly spherical regions of radius 200 \h Mpc, whereas our actual peculiar velocity  corrections are based off the 2M++ survey which has two imposed flux limits (and corresponding distance cuts) imposed resulting in an effective volume equivalent to a sphere with a radius of 176.5 \h Mpc.

    \item  We find that the density fluctuations of the particle density field measured within these 200 \h Mpc spheres ($0.0178 \pm 0.003$) does not fully align with predictions of \LCDM\ matter fluctuations (0.0247).
    This indicates that there is a potential discrepancy with regards to the bias for the treated SN objects. This indicates that we may be underestimating the effect of $\Delta\sbr{g}$ in our measurement of eq. (\ref{eqn:v_pred}). To account for this potential bias we choose to introduce the ratio between the \LCDM\ predictions to the MDPL2 particles, $\varkappa=1.39$,    
    into our final measurements to ensure that we are not underestimating our measurements of $\sigma_{H_0}$. This is done in the following manner; given a set number of SN the following is true:
    \begin{equation}
\log(\Ho(z))) - \log(\Ho(\zcosmo)) =\sum_{i} \left[\log(\Ho(z_i)) - \log(\Ho(\zcosmo))\right]  \\,
    \end{equation}
    here $i$ denotes the condition of each set that makes up the full set of SN redshifts. For the simplest case these would be:
    \begin{itemize}
        \item \Ho($z_{<2M++}$): changes in redshift are only applied to objects within 2M++, anything outside of the survey volume is kept at \zcosmo.
        \item \Ho($z_{>2M++}$): Changes in redshift are applied solely to objects outside of 2M++.
    \end{itemize}
    The sum is hence a combination representing the various components that make up our final $z$.  The uncertainty in \Ho($z$) is thus composed of the uncertainties of the individual \Ho($z_i$) components. Given that these are also correlated, we can break the measurement of  $\sigma_{H_0}(z)$ into a combination of the individual $\sigma_{H_0}(z_i)$ such that:
    \begin{equation}
        \sigma_{H_0}(z) = \sqrt{ \sum_i \alpha_i^2\sigma^2_{H_0}(z_i) +\sum_i \sum_{j\neq i} \alpha_i\alpha_j\sigma_{H_0}(z_i)\sigma_{H_0}(z_j)\rho_{i,j}}   \\,
    \end{equation}
    here $\alpha_i =\pm1, \varkappa$ representing the condition of the how we are combining these uncertainties. The correlation coefficient between measurements of \Ho(z$_i$) and  \Ho(z$_j$) is given by:
    \begin{equation}
        \rho_{i,j} = \frac{\left< \varepsilon_{H_0}(z_i)  \varepsilon_{H_0}(z_i)\right>}{ \sigma_{H_0}(z_j)  \sigma_{H_0}(z_j)} \\,
    \end{equation}
    where $\varepsilon_x = x- \left< x \right>$.
 
    Our final quoted $\sigma_{H_0}$(\zcmb) and $\sigma_{H_0}$(\zrec) measurements in table \ref{table:HOcomp}\ are adjusted as follows, for the components of $\sigma_{a_x(z)}$ within 2M++ we subtract off in quadrature the contributions that come from the reconstructed peculiar velocities before adding them back accounting for $\varkappa$. 
    We do not perform any corrections to \Ho\ measurements beyond 2M++.
\end{itemize}

\subsection{Results}\label{sec:results}

Figure \ref{fig:axcomp} shows the histograms of the differences between recovered \Ho\ measurements of the `observed' CMB frame redshifts (solid) and the reconstructed redshift measurements (hatched) from the mocks' true cosmological redshift measurements of \Ho. As demonstrated here for the SH0ES HF sample, the reconstruction reduces the uncertainty in the measurement of \Ho\ such that $\sigma_{\Ho}(\zrec)=0.31$
\kmsmpc, which is a factor of $\sim 1.5$ reduction when compared to the case where \zcmb\ is used.

Of course this sample consists of a very limited number of SNe, with a not insignificant number lying beyond the survey limits of 2M++.  In table \ref{table:HOcomp} we show our r.m.s. \Ho\ errors for several  redshift ranges that have been considered in the literature.  The top portion of this table shows our measurements which include the calculations performed using the redshifts modified using the constrained fluctuations of objects that lie outside of 2M++. While the bottom portion of the table explores the same minimum redshift cuts, but here we limit our sample to SNe lying within 2M++.  We find that, regardless of imposed minimum redshift cuts, all measurements with the same maximum redshift of $\sigma_{H_0}(\zrec)$ are very similar. 
However, unsurprisingly, \Ho\ measurements based on the CMB redshifts are more heavily impacted at lower redshifts due to the (uncorrected) peculiar velocities. We additionally find that using no peculiar velocity corrections in the recovered \Ho, leads to significant uncertainty: $\sigma_{H_0}(z\sbr{CMB})\sim 0.4-0.8$ \kmsmpc\ (0.5-1.1 \kmsmpc\ if we include reconstructed redshifts of all objects including calibrators to vary).  However, we find that the reconstructions perform comparably well for all explored redshifts samples. 
In all cases we find that the reconstruction only slightly impact our recovered value of $H_0$ by a maximum of $\sim 0.4$ \kmsmpc. 

We additionally list here the effects of both the unconstrained monopole density fluctuations of section \ref{sec:monopole}, compared to the 2M++ constrained density fluctuations.

\begin{table}[!ht]\label{table:HOcomp}
    \centering
    \begin{tabular}{l c c  c c c c }
    \hline
        redshift range & N  & $\sigma_{H_0}(\delta_{\rm uncon})$ & $\sigma_{H_0}(\delta_{\rm con})$ &  $\sigma_{H_0}(\zcmb)$ & $\sigma_{H_0}(\zrec)$ \\ \hline
\textbf{SH0ES HF SN}  & \textbf{277} & \textbf{0.57} $\pm$ \textbf{0.02} & \textbf{0.49} $\pm$ \textbf{0.02} & \textbf{0.48} & \textbf{0.31} \\
Full SN sample & 1624 &  0.44 $\pm$ 0.01 & 0.42 $\pm$ 0.01   & 0.40 & 0.18 \\
$0.01<z<0.15$ & 705 &  0.65 $\pm$ 0.02 & 0.58 $\pm$ 0.02   & 0.56 & 0.29 \\
$0.0233<z<0.15$ & 490 & 0.53 $\pm$ 0.02 & 0.45 $\pm$ 0.02   & 0.43 & 0.27 \\ \hline
SH0ES HF SN in 2M++ & 203 & 0.81 $\pm$ 0.03 & 0.72 $\pm$ 0.02  & 0.69 & 0.42 \\ 
all 2M++ & 601 & 0.95 $\pm$ 0.03 & 0.87 $\pm$ 0.03  & 0.84 & 0.41 \\
$0.01<z<$ 2M++ & 566 & 0.94 $\pm$ 0.03 & 0.86 $\pm$ 0.03  & 0.81 & 0.42 \\
$0.0233<z<$ 2M++ & 338 & 0.80 $\pm$ 0.03  & 0.71 $\pm$ 0.02 & 0.68 & 0.41\\
    \end{tabular}
    \caption{
Column (1) the sample and redshift range and (2) the number of objects used in the used in the calculation of the standard deviation of \Ho.
The following four columns show the ``cosmic variance'' standard deviation in the error in \Ho, using 4 different methods.
This calculation is done with respect to measurements using the `true' cosmological redshift and: (3,4) the unconstrained, 200 \h Mpc constrained and the 2M++ constrained velocity fluctuations (see section \ref{sec:monopole}); (5) The `CMB' frame redshift; and (6) the reconstructed redshift outlined in section \ref{sec:recon}.  }
\end{table}

\section{Discussion}\label{sec:discussion}

We used an analytic method to measure the systematic error in  measurements of the \Ho, due to local  large-scale density fluctuations. 
Matching \LCDM\ predictions and using the full Pantheon+ SN sample, we find this impacts the recovered value of \Ho\ by $0.44 \pm 0.01$ \kmsmpc. For the HF SN sample, this effect is slightly larger at $0.57 \pm 0.02$ \kmsmpc. While both of these are larger than the measured density fluctuations of our mocks in our CMB redshift measurements of 0.4, 0.48 respectively, they are comparable to the variations of our constrained fluctuations. We attribute this discrepancy to our limited number of mocks. Thus we estimate that, without predicted peculiar velocity corrections, cosmic variance could impact the R22 estimate of \Ho\ by 0.6\%.  However, when incorporating redshift reconstruction this impact is reduced to 0.24\%, a factor of $\sim$40 less than the 9.0\% change required for the measurement to reconcile with the findings of ref. \cite{planck_collaboration_planck_2020}.

We now discuss whether there are additional sources of uncertainties which are not captured in our methodology which could significantly increase the variance estimated in Section \ref{sec:results}.

\subsection{Reconstruction using Calibrator Objects}\label{sec:calibrators}

Throughout this work, we have considered the objects classified as calibrators in Pantheon+ as non-existent when calculating the effects of reconstruction on measurements of \Ho. We adopt this methodology for the following reasons. 
\begin{itemize}
    \item   The measurements of SNe at higher redshifts rely heavily on those from lower redshift calibrators. The inclusion of these low-redshift objects in our calculations, could introduce biases or inaccuracies as any errors in the calibration redshifts propagate into higher redshift measurements.
    \item Peculiar velocities and reconstruction can significantly impact measurements for these very low redshift objects, which span from $0.001 < z < 0.017$ or approximately 350 - 5000 km/s, and thus any error in the velocity reconstruction heavily impacts the objects final redshift.
\end{itemize}
If we repeat our estimation of \Ho\ variation using \emph{only} the calibrator sample, we find variations in \Ho\ on the order of 2.0 \kmsmpc\ due to peculiar velocity redshift reconstruction. To justify this further we calculated the change in the recovered values of \Ho\ including calibrators using the following methodologies: (1) all calibrators objects are kept fixed at the redshifts provided in the Pantheon+ catalogues, all other objects' redshifts are reconstructed, (2) all objects in the Pantheon+ catalogue undergo redshift reconstruction, (3) calibrators are omitted from our calculations (the default scheme adopted in this paper). 
As shown in table \ref{table:cals} including calibrators and their redshift to vary in our \Ho\ calculations results in an increase of 9\% for our 2M++ sample and 4\% in the full Pantheon+ measurements. This highlights the impact of calibrator redshifts on the final measurement of \Ho\, but as it is not the focus of this paper we simply ignore these objects in our calculations.

\begin{table}[!ht]\label{table:cals}
    \centering
    \begin{tabular}{l c c c }
    \hline
        redshift range &  reconstructed $z_{cal}$ & fixed $z_{cal}$ & omitting  $z_{cal}$  \\ \hline
Full SN sample &  0.18 & 0.17 & 0.18 \\
all 2M++ & 0.42 & 0.39 & 0.41 \\

    \end{tabular}
    \caption{
Column (1) the sample and redshift range. The following three columns show the standard deviation in the measurement of \Ho\ in \kmsmpc, using three different methods to account for calibrators in the sample.
This calculation is done with respect to measurements using the `true' cosmological redshift and the reconstructed redshift: (2) all calibrator redshifts are reconstructed using the method outlined in Sec \ref{sec:recon}, (3) all calibrators are kept at their given  Pantheon+ redshifts, (4) all calibrators are omitted in the calculation.}
\end{table}

\subsection{Beyond 2M++}\label{sec:calibrators}

We note that the R22 analysis does not correct for peculiar velocities outside of the survey volume, but use measurements comparable to \zcmb\ measurements to within $\sim 75$ km s$^{-1}$. Using our density fluctuation method as peculiar velocity proxies, we estimate that this generates a negligible change in the measurement of \Ho\ of $0.02$ \kmsmpc\ for the full SN sample. This is a small deviation as expected, in part due to the fact that at higher redshift the fractional uncertainty in $\Delta z/ z$ is small as the redshift measurements uncertainties dominate over the those of the peculiar velocities.  

\subsection{Classification of Centrals}\label{sec:calibrators}

We additionally evaluate the consequences of incorrectly identifying the central galaxy in our analysis. The reconstruction of the 2M++ density and velocity field, is partially dependent on identifying the catalogue's central galaxies. We have only been using objects classified as central galaxies in our simulations for this purpose. However, if we use the methodology outlined in ref. \cite{lavaux_2m_2011} to identify central galaxies and find their best redshift, when using these objects as the base for our calculation the differences in the overall measurement of the \Ho\ is negligible impacting measurements of \Ho\ for the HF sample by 0.01 \kmsmpc.  Therefore as expected, the impact of not correctly identifying the central galaxy appears to be subdominant in relation to the overall impact on velocity reconstructions.

\section{Conclusions}\label{sec:conclusions}

By using mock 2M++ density fields that mimic the selection effects of the reconstructed density field used in recent \Ho\ SNe analyses, we are able, for the very first time, to directly measure the uncertainty introduced in measurements of the Hubble constant due to reconstruction errors. We modelled scatter in cosmological redshifts beyond the survey limits of 2M++ by introducing velocity fluctuations from the density field variations. We find that our reconstruction method is able to recover measurements of \Ho\ to within $\sim$0.4 \kmsmpc\ for \Ho\ measurements solely within 2M++ or 0.2 \kmsmpc\ for the full Pantheon+ SN range. In either case we find reconstruction is able to improve the upon CMB-frame measurements by a factor of $\sim 1.5-2$ in \Ho. 

Several analyses of \Ho\ have omitted very low redshift SN laying because they were not considered to be part of the Hubble Flow. However, we find that extending the SN sample to a redshift of 0, we find no discernible difference in the standard deviation of the recovered \Ho\ measurements. This shows that the 2M++ peculiar velocity corrections are effective at these low redshifts. It should therefore be possible to include these objects with appropriate weights, as the scatter in the recovered $a_x$ is only minimally affected by these low-redshift data. This is left for future work.

As the density field is normalised within the survey volume of 2M++, the peculiar velocity predictions generated using this field would not account for any local under-densities on the scale of 200 \h Mpc or larger. Any such large-scale local density fluctuation would manifest as a systematic error in the predicted velocities compared to the true velocities. Thus our mock catalogues are only able to capture this effect on 2M++ scales, and effects due to larger scales are accounted for using our constrained fluctuation method. %

We find that the uncertainty in $H_0$ arising from peculiar velocities in the Hubble flow is estimated to be $\sim 0.3$ \kmsmpc, hence can only account for a small part in the overall uncertainty and cannot appreciably alleviate the $H_0$ tension.

\acknowledgments

AMH and MJH acknowledge support from the Natural Sciences and Engineering Research Council Discovery grants programme and from the University of Waterloo.

This research was enabled in part by support provided by Compute Ontario (computeontario.ca) and by the Digital Research Alliance of Canada (alliance.can.ca)

The \texttt{CosmoSim} database used in this paper is a service by the Leibniz-Institute for Astrophysics Potsdam (AIP). The MultiDark database was developed in cooperation with the Spanish MultiDark Consolider Project CSD2009-00064. The authors gratefully acknowledge the Gauss Centre for Supercomputing e.V. (www.gauss-centre.eu) and the Partnership for Advanced Supercomputing in Europe (PRACE, www.prace-ri.eu) for funding the MultiDark simulation project by providing computing time on the GCS Supercomputer SuperMUC at Leibniz Supercomputing Centre (LRZ, www.lrz.de).




\bibliographystyle{jhep}
\bibliography{PVs}


\end{document}